# Hierarchical Bayesian Model for Probabilistic Analysis of Electric Vehicle Battery Degradation

Mehdi Jafari, *Member, IEEE*, Laura E. Brown, and Lucia Gauchia, *Senior Member, IEEE*

*Abstract*— This paper proposes a hierarchical Bayesian model for probabilistic estimation of the electric vehicle battery capacity fade. Since the battery aging factors such as temperature, current, and state of charge are not fixed, and they change in different times, locations and by the different users, deterministic models with constant parameters cannot accurately evaluate the battery capacity fade. Therefore, a probabilistic presentation of the capacity fade including uncertainties of the measurements/ observations of the variables can be a proper solution. We have developed a hierarchical Bayesian Network model for the electric vehicle battery capacity fade considering multiple external variables. The mathematical expression of the model is extracted based on Bayes' theorem, the probability distributions for all variables and their dependencies are carefully chosen where the Metropolis-Hastings Markov Chain Monte Carlo sampling method is applied to generate the posterior distributions. The model is trained with a subset of experimental data (85%) to obtain its unseen parameters and tested with other 15% of data to prove its accuracy. Also, three case studies for different drivers, different grid services' frequencies, and different climates are explored to show model's flexibility with different input data. The developed model needs training data for parameter tuning in different conditions. However, after training, it has more than 95% precision in estimating the battery capacity fade percentage.

## I. Introduction

BATTERY degradation is evaluated by two main approaches: data-driven prognostics and model-based methods [1]. The data-driven methods are mostly used for real-time applications and after training through a set of data, they use present measurement data from the battery to predict the battery state of health (SOH). Different types of Kalman filters (KF) [2], relevance vector machine (RVM) [3] and functional principal component analyses (FPCA) [4] are some of these methods. These methods need expert knowledge to relate the available data to the battery health condition. However, they are less computationally intensive and are easy to implement in online applications.

The model-based approach, on the other hand, is based on the historical data of the battery performance and relies on the mathematical expression for the battery capacity fade evaluation [5]. Electrochemical models, equivalent circuit models (ECMs), and empirical aging models are different groups of this approach. Electrochemical models evaluate the battery degradation with chemical properties of the battery components such as solid electrolyte interface (SEI) layer growth, cycleable lithium loss or lithium concentration decrease [6], [7]. ECMs simulate the battery aging by varying the circuit components' values with cycling and mainly use the electrochemical impedance spectroscopy (EIS) test results to obtain them [8], [9]. The empirical aging models [10]–[12] fit mathematical equations to extensive experimental results obtained from accelerated aging tests on the battery. These models mostly use different forms of the Arrhenius equation to include the effect of different aging factors such as state of charge (SOC), C-rate and temperature.

The shortcoming of the model-based aging studies is that they rely on deterministic mathematical equations and consider the aging effectual factors as constant values during cycling. However, neither the aging phenomena is deterministic, nor its effecting factors stay constant during the life of the battery. In an electric vehicle (EV) application, the capacity fade of the battery during cycling is directly affected by the driver's behavior, charging facilities availability, grid services presence, and the temperature/weather conditions [13]. None of these factors can be defined by fixed values and they carry significant uncertainties. Therefore, deterministic models are not reliable in evaluating the battery capacity fade for different usage conditions and probabilistic methods including the uncertainties of the measurements and processes such as Bayesian models can present more informative and accurate evaluation of the battery capacity fade [14].

Bayesian models are applied to estimate the battery SOH, which are mostly proposed for online applications. He et al. [15] proposed the online SOH estimation with Dynamic Bayesian networks (DBN). In this study, the terminal voltage measurements in the constant current charging cycle are used to estimate the state of charge (SOC) of the battery in a different time interval.

This work was supported by the National Science Foundation through the project CAREER: An Ecologically Inspired Approach to Battery Lifetime Analysis and Testing, under Award 1651256.

M. Jafari was with Department of Electrical and Computer Engineering, Michigan Technological University, Houghton, MI, USA. He is now with Massachusetts Institute of Technology (MIT) LIDS, 32 Vassar St. Cambridge, MA, USA (e-mail: jafari@mit.edu).

Laura E. Brown is with Department of Computer Science, Michigan Technological University, Houghton, MI, USA (e-mail: lebrown@mtu.edu).

Lucia Gauchia is the Department of Electrical and Computer Engineering, and Department of Mechanical Engineering-Engineering Mechanics, Michigan Technological University, Houghton, MI, USA (e-mail: gauchia@mtu.edu).

**Nomenclature:**

*Observations (measurements):*

| | |
|---|---|
| q | Capacity fade percentage |
| T | Temperature |
| a | Vehicle acceleration |
| v | Vehicle Velocity |

*Unobserved variables:*

*Battery specific:*

| | |
|---|---|
| $\lambda$ | Estimated capacity fade percentage |
| $\hat{T}$ | Estimated temperature |
| $Ah$ | Ampere-hour |
| $SOC$ | State of charge |
| $I_c$ | Battery current (C-rates) |
| $u$ | Battery voltage |
| $P_b$ | Battery power |
| $\delta$ | Charging/discharging efficiency |

*Vehicle and grid specific:*

| | |
|---|---|
| $P_w$ | Wheel power |
| $P_g$ | Grid power (charge, discharge and grid services) |
| $P_{aux}$ | Vehicle auxiliary power |
| $\hat{a}$ | Estimated acceleration |
| $\hat{v}$ | Estimated velocity |

*Parameters:*

| | |
|---|---|
| $\alpha$ | Aging linear dependency to SOC: slope |
| $\beta$ | Aging linear dependency to SOC: y-intercept |
| $\varepsilon$ | Residual error of capacity fade estimation |
| $\zeta$ | Power factor for Ah dependency |
| $\eta$ | C-rate factor |
| $E_a$ | Activation energy |
| $\gamma$ | Estimated total vehicle mass including passengers |
| $\omega$ | Vehicle aerodynamic drag |
| $\varphi$ | Estimated road grade |
| $k_1$ | Driving contribution coefficient |
| $k_2$ | L1 charging coefficient |
| $k_3$ | L2 charging coefficient |
| $k_4$ | L3 fast DC charging coefficient |
| $k_5$ | Solar integration coefficient |
| $k_6$ | Frequency regulations contribution coefficient |
| $k_7$ | Peak shaving Ah demand coefficient |

*Indices:*

| | |
|---|---|
| $i$ | Observations $i=1,\ldots,N$ |
| $j$ | Capacity fade percentage measurements $j=1,\ldots,J$ |
| $m$ | Temperature measurements $m=1,\ldots,M$ |
| $k$ | Acceleration and velocity measurements $k=1,\ldots,K$ |

*Deterministic functions:*

| | |
|---|---|
| $f$ | Capacity fade variables relationship |
| $g$ | Battery power calculation |
| $y$ | Battery current calculation |
| $h$ | Vehicle mechanical power relationship |

The model uses the charging voltage profile as input and gives the probability of the battery being in different capacity classes. Similarly, Jin et al. have estimated the spacecraft secondary life battery's capacity fade using online charge/discharge measurements [16]. In [17], Electrochemical Impedance Spectroscopy (EIS) and capacity test results are used as observations to develop a Bayesian network (BN) and estimate the battery SOH with internal impedance and capacity. A simplified Bayesian model known as "naive" Bayes is proposed in [18] to predict the battery SOH which eliminates the hierarchical parts of Bayesian network which present the intermediate hidden states. Mishra et. al. [19] have maintained the hierarchical properties of BN in their model, however, they use a simple ECM model and estimate the end of life of the battery based on discharge measurements and do not consider the effect of temperature and dynamics of the cycling current. In [19] and [20], capacity measurements were used to make prediction of the remaining useful life (RUL) of the battery. In these previous Bayesian approaches, there is a more focused effort on using it as a classifier or estimator without considering the external factors and their variations. In addition, these models are developed for online applications with real-time measurements of battery's electrical signals.

Bayesian models can also be applied to probabilistic analysis of phenomena which deal with multiple external stochastic factors and unmeasurable variables. This application of Bayesian models is widely used in ecological studies and proved to be a successful probabilistic analysis tool for the natural and physical systems [22], [23]. Considering the large amount of available data for the EV driving, recharging and grid services such as solar charging which contains uncertainties and measurement errors, and their hierarchical effect on the battery aging, this application of Bayesian models can be useful for the aging probabilistic analysis. Causality is of utmost importance for batteries as their aging is affected by a high number of hierarchical variables that depend upon external factors to the battery.

Therefore, acknowledging the advantage of Bayesian models in considering the uncertainties of the variables and providing probability distributions instead of point value estimations, we propose a hierarchical Bayesian model for the probabilistic battery capacity fade evaluation in the EV application. The initial idea of this work is presented in [24] which includes a basic aging model to study the effect of driving on the capacity fade. In this paper, more detailed Bayesian model is developed which has three times more variables to include all possible sources of uncertainties and it is used to evaluate the effects charging and grid services variation on the battery degradation, besides driving. Through more precise degradation estimations, this model can help manufacturers in optimum warranty definition and the drivers by onboard battery health evaluation. We have focused the paper around proposing a Bayesian framework that can relate external factors to battery aging evaluation. The main contributions of this work are: proposing an ecologically-inspired BN analysis which covers causality and hierarchy in the EV battery aging phenomena in a way that can consider multiple causes and scales; incorporating a variety of aging external factors from driving style and EV's electrical/mechanical variables to grid interface into the model; and probabilistic analysis and quantifying the EV battery capacity fade in the case studies with real-world data. The Bayesian network for this problem is developed including all observed and unobserved variables and mathematical presentation of the model is extracted. The probability distributions for the variables are defined and Markov Chain Monte Carlo (MCMC) sampling is employed to obtain the posterior distributions. The model training, test and case study are presented to show its performance. This model reflects the uncertainties of measurements and process, it precisely evaluates the capacity fade and provides more informative results, and it is applicable for any type of input data with proper training.

## II. BAYESIAN MODELS AND EV BATTERY AGING FACTORS

Contrary to deterministic models that use direct mathematical equations to present the relationship between variables, a BN consists of two parts <G, P> a directed acyclic graph G and parameter P to determine a joint probability distribution over the random variables of the network. The BN satisfies the Markov condition, every variable is independent of any subset of its non-descendants conditioned on its parents [25]. Graphical representation of the BN includes "nodes" and "edges" in which each node is a random variable connected to its parent nodes through edges. A key point of BN is that it classifies the variables into two groups: "Observations" and "Unobserved variables" and treats all unobserved quantities as random variables. Based on the Bayes' theorem, the unobserved variables including unmeasured/unseen quantities and model parameters (**x**) are conditional to observations/measurements (**y**) as follow:

$$P(\mathbf{x}|\mathbf{y}) = P(\mathbf{y}|\mathbf{x})P(\mathbf{x})/P(\mathbf{y}) \quad (1)$$

where, $P(\mathbf{x}|\mathbf{y})$ is the posterior distribution, $P(\mathbf{y}|\mathbf{x})$ is the likelihood of observations, $P(\mathbf{x})$ is the prior distribution of unobserved variables, and $P(\mathbf{y})$ is the marginal distribution of vector of observations and can be calculated by (2). The bold style denotes "vector" of variables.

$$P(\mathbf{y}) = \int P(\mathbf{y}|\mathbf{x})P(\mathbf{x})d\mathbf{x}. \quad (2)$$

This presentation of a problem is capable of including all uncertainties of the measurements, sampling, parameters, and modeling. However, analytically solving (2) is not possible for large number of the unobserved variables **x**. Therefore, it should be solved by numerical approximation methods like Markov Chain Monte Carlo (MCMC) sampling.

The main capacity fade factors of the electric vehicle battery are temperature, state of charge (SOC), cycling C-rate, and total Ah throughput (3) [10], [26]. Among these factors, temperature is independent, and all other factors are dependent to the frequency of daily tasks (driving, recharging, utility services) and human factors (driving styles and recharging habits).

$$Q_{loss} = f(T, SOC, I_C, Ah). \quad (3)$$

None of these factors are fully determined and there are uncertainties in each. Temperature patterns cannot be fully predicted, the driving distance and style are dependent to the needs of the vehicle owner and his/her personality, and recharging and possibility of the utility services are upon access to the power grid. Therefore, it is not possible to accurately estimate the battery capacity fade using a deterministic mathematical model while its variables are stochastic, and BN probabilistic approach can be a proper solution for the battery capacity fade modeling.

## III. HIERARCHICAL BAYESIAN MODELING PROCESS

### A. Network Development

The developed Bayesian network in [24] only considers the "aging variables" with total of 14 nodes. In this work, the network is enhanced to be more complete and includes the EV's "electrical and mechanical variable" as well as the "charging and grid interface variables" with total of 39 nodes to present all tasks of an EV battery. The following explanations illustrates the developed BN and the hierarchy in its variables dependency, which will help to construct the Bayesian model. In the first step to develop the Bayesian model, observations and unobserved quantities are defined. The battery cell's capacity fade percentage is measurable ($q$), however it has measurement uncertainty and it should be estimated to consider the measurement error ($\lambda$). The estimated capacity fade percentage is affected by the total Ampere-hour throughput ($Ah$), state of charge ($SOC$), temperature estimation ($\hat{T}$), and C-rates ($I_c$). It is important to note that the factor of time which is one of the main effectual factors in the battery life is not directly considered in the formulation, however it is accounted for indirectly by the factor of total Ah throughput [10], [27]. Considering the battery's daily tasks and as result of cycling in those tasks, total Ah throughput is limited to the time. The temperature estimation can be obtained from its measurement data ($T$) including the measurement and sampling uncertainties. It is considered that the observation data have measurement and sampling errors (except if it is perfectly observed) and to include that uncertainty in the results, an "estimation" intermediate variable is defined which estimates the probability function of those data to reflect their uncertainties. SOC and Ah can be obtained from C-rates, and C-rates are dependent on the battery voltage ($u$) and battery power ($P_b$). The battery voltage has a distribution between its fully charged and fully discharged voltage values. $P_b$, on the other hand, is dependent to the wheel power ($P_w$), auxiliary power ($P_{aux}$), EV mechanical and electrical parts' efficiencies index ($\delta$), the charging/grid services power ($P_g$) and all these factors' contribution coefficients ($k_n$, $n=1,2...,7$). The contribution coefficients are considered for probabilities of driving, standard L1, L2, and L3 chargers, frequency regulations, peak shaving and solar power integration. The powers for charging and grid services ($P_{L1-3}$, $P_s$, $P_f$, $P_{ps}$) are considered to be perfectly observed and therefore they do not reflect any uncertainty. Also, $P_{aux}$ is considered to have a distribution between 5-15% of the driving power. The charging power is assumed to be known and the grid services is performed by a standard L2 charging station.

The wheel power can be calculated from the vehicle velocity estimation ($\hat{v}$) and the acceleration estimation ($\hat{a}$) using the governing mechanical equations. $\hat{v}$ and $\hat{a}$ are obtained from the velocity and acceleration time series ($v$ and $a$) considering their measurement and sampling errors. Also, parameters are defined to relate all these variables dependencies as aging parameters, $\alpha$, $\beta$, $E_a$, $\eta$, $\zeta$, $\varepsilon$, and vehicle mechanical parameters, $\gamma$, $\omega$, and $\varphi$. In our problem, there are $i=1...N$ observations and in each observation, we have $K$, $M$, and $J$ measurements for $v$ and $a$, $T$, and $q$, respectively.

After defining the causality in the aging model and different variables' dependency, the Bayesian network can be designed out of the defined variable. Fig. 1 can be used to classify the observations, variables and parameters to generate the sketch of BN. Using this diagram and our expert domain knowledge we have created the network structure as shown in Fig. 2. The key point in creating the BN structure is to define the parents-children node relationship among the variables. For the observations (e.g. velocity data $v$), a parent node of estimation should be defined ($\hat{v}$). Then, the defined parent node along with other parent nodes can be used to define a children node (e.g. wheel power $P_w$). Definition of relationship between parents and children nodes should follow a solid logic that can be either a known equation or a statistical data assessment with parameters. To account for unmeasurable factors and consider their uncertainties, extra parameters can be defined and calculated in the model training. Note that some nodes can represent the perfectly measured quantities that do not have uncertainties.

Other such models may be differently developed but statistically indistinguishable based on the independencies/dependencies entailed by the model. The network has three main parts: "data" shows the observations/measurements, "process model" includes the aging and EV variables and "parameters" define the model variables' dependencies. Solid edges refer to the probabilistic relationship, while the dashed edges show deterministic dependencies and the nodes at the beginning of these arrows are not random variables. It is important to note that while defining the variables (network nodes) and their dependency, it is not possible to include all the effectual factors on the network variables and some causality may be missing in the network development. However, the merit of Bayesian model is that it considers the uncertainties associated with the factors that are not directly connected in the network structure by assigning an uncertainty parameter for each variable. Therefore, presenting each node with probability distribution calculated based on the rest of network, it considers a degree of uncertainty to reflect the effects that are not directly addressed.

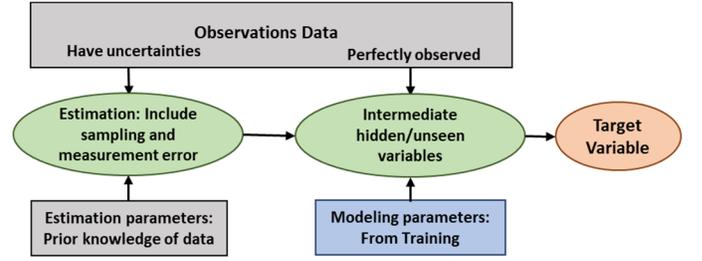

Fig. 1. Bayesian network design diagram

### B. Mathematical Expression

Considering the nodes and edges of the network in Fig. 1, the mathematical expression of the model considering all variables' uncertainties can be written as (4) to (8). In (4), "[ ]" denotes the "probability of" the variable inside the brackets. Note that the sign between left and right side of the equation is not "=", it is "∝", because the marginal distributions of the observations (refer to the denominator of (1)) are not included in this equation. Therefore, the MCMC sampling method is used to draw large number of samples from the posterior distributions and estimate them with distribution fitting. To apply MCMC, the probability distribution function (PDF) of all variables should be defined based on their properties and full conditionals for all variables should be written.

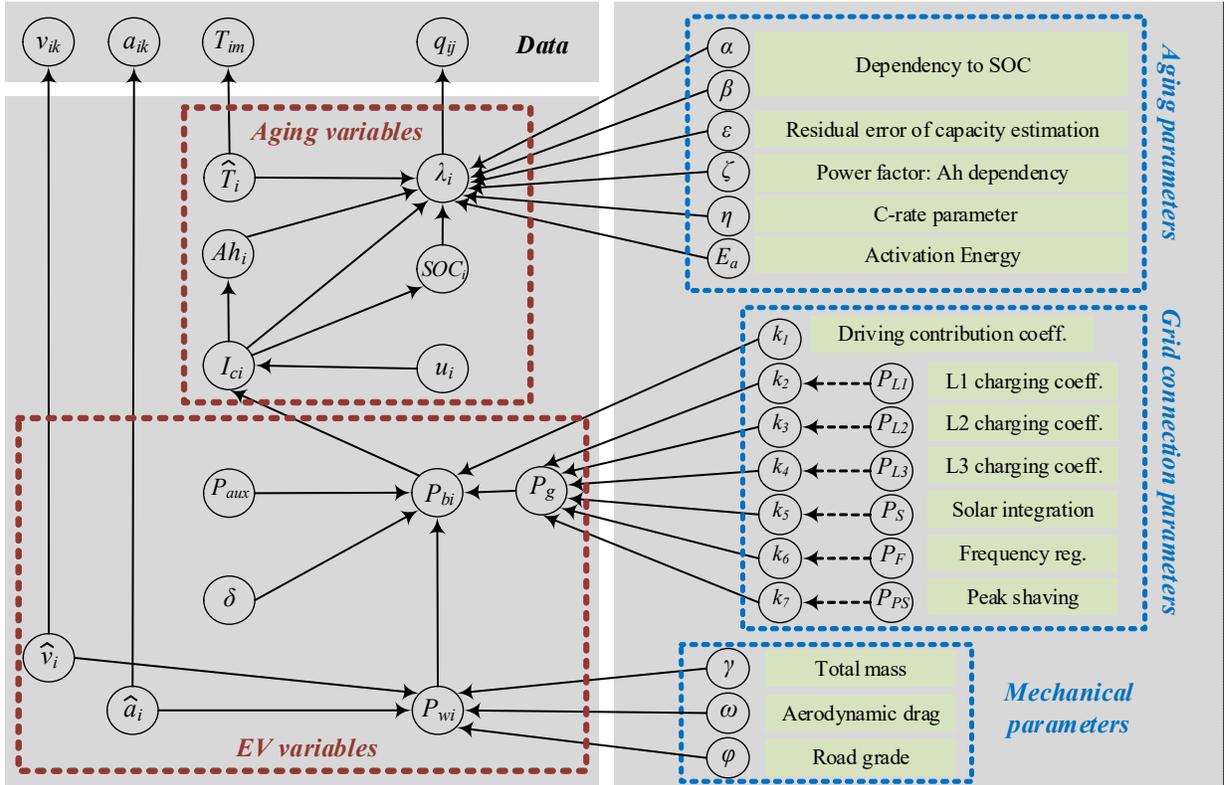

Fig. 2. Graphical view of the developed BN

$$[\lambda, \hat{T}, I_c, Ah, SOC, u, P_b, P_w, \hat{a}, \hat{v}, P_{aux}, \delta, \alpha, \beta, E_a, \eta, \zeta, \varepsilon, k_n, \gamma, \omega, \varphi | \mathbf{a}, \mathbf{v}, \mathbf{T}, \mathbf{q}] \propto$$

$$\prod_{i=1}^{N}\prod_{j=1}^{J}[q_{ij}|\lambda_i] \times [\lambda_i|f_i] \times \prod_{m=1}^{M}[T_{im}|\hat{T}_i] \times [\hat{T}_i] \times \prod_{k=1}^{K}[a_{ik}|\hat{a}_i] \times [\hat{a}_i] \times [v_{ik}|\hat{v}_i] \times [\hat{v}_i] \quad (4)$$

$$\times [Ah_i|I_{ci}] \times [SOC_i|I_{ci}] \times [I_{ci}|y_i] \times [u_i] \times [P_{bi}|g_i] \times [P_{aux}] \times [\delta] \times [P_{wi}|h_i]$$

$$\times [\alpha] \times [\beta] \times [E_a] \times [\eta] \times [\zeta] \times [\varepsilon] \times [k_n] \times [\gamma] \times [\omega] \times [\varphi]$$

$$f_i = (\alpha SOC_i + \beta)\exp\left(\frac{-(E_a - \eta I_{ci})}{R\hat{T}_i}\right)Ah_i^{\zeta} + \varepsilon \quad (5)$$

$$y_i = \frac{P_{bi}}{u_i} \quad (6)$$

$$g_i = k_1\left(\frac{1}{\delta}P_{wi} + P_{aux}\right) + k_n P_{gn} \quad n = 2,\dots,7 \quad (7)$$

$$h_i = \gamma \hat{a}_i \hat{v}_i + \omega \hat{v}_i^3 + \varphi \hat{v}_i \quad (8)$$

Referring to the model's variables and exploring their properties, it is possible to define a proper distribution function for them. Among variables, λ, *Ah*, *u*, and $\bar{v}$ are definitely equal to or greater than zero. Therefore, a gamma PDF is used to define their distribution. For the same reason, the gamma PDF is used to define the parameters, too. The *SOC* and *δ* values vary in [0-1] interval, so a Beta PDF is more appropriate for these variables. C-rates are mostly concentrated on positive small values and it has a tail toward larger numbers. Therefore, a Rayleigh PDF defines its properties better. The rest of variables including powers, acceleration and temperature are defined with the normal PDF as they are centered around a value with possibility of both positive and negative tails. Using the defined PDFs, we have derived the full conditionals for all variables to apply the MCMC. An example of full conditionals for λ, $I_c$, $\bar{v}$, and δ are presented in (9) to (12).

$$[\lambda|\cdot] \propto \prod_{i=1}^{N}\prod_{j=1}^{J}N(q_{ij}|\lambda_i) \times G(\lambda_i|f_i) \quad (9)$$

$$[I_c|\cdot] \propto \prod_{i=1}^{N}G(\lambda_i|f_i) \times G(Ah_i|I_{ci}) \quad (10)$$
$$\times B(SOC_i|I_{ci}) \times R(I_{ci}|y_i)$$

$$[\hat{v}|\cdot] \propto \prod_{i=1}^{N}N(P_{wi}|h_i)\prod_{k=1}^{K}G(v_{ik}|\hat{v}_i) \times G(\hat{v}_i) \quad (11)$$

$$[\delta|\cdot] \propto \prod_{i=1}^{N}N(P_{bi}|g_i)B(\delta) \quad (12)$$

[λ|·] stands for the λ conditional to all the variables related to it and it should include all the terms that have λ from (4). Also, N, G, B, and R refer to normal, gamma, beta, and Rayleigh PDFs, respectively. It is important to note that all these PDFs require values for their standard deviation (SD) to calculate the probabilities, which are defined based on the properties of each variable.

### C. MCMC Sampling Implementation

The model requires the initial values for all variables to initiate the MCMC. Note that, because the model calculates 10,000 Monte Carlo samples based on the conditional probabilities to the observations, and the estimation is made out of all samples, the first sample which is the initial guess does not play a vital role. For the unseen variables such as *SOC*, *Ah* and powers, calculations from the input data are used to define the initial guess. However, for the parameters which are more important to be initiated properly, different sources of information are used. *ε* is the residual of (5) and therefore, its initial value is considered to be zero. $E_a$ and *ζ* are activation energy and power factor respectively, and based on the reports from literature [27], their initial guess is 31000 and 0.5, respectively. Initial values for *α, β,* and *η* are selected based on the engineering judgment. These values are selected in a way that the capacity fade estimation on that single sample is in the range of observations. Considering the governing mechanical equations of the vehicle, *γ* reflects the total vehicle mass which is the mass of vehicle in addition to the mass of passengers. Initial value of *ω* is calculated from (13) to reflect the effect of aerodynamic drag of the vehicle and for *φ*, the initial value is calculated by (14) to include the impact of rolling resistance and road grade.

$$\omega^{(1)} = 0.5\rho C_x A_f \quad (13)$$
$$\varphi^{(1)} = mgf_r\cos\theta + mg\sin\theta \quad (14)$$

where, *ρ* is the air density, $C_x$ is the aerodynamic coefficient, $A_f$ is the vehicle frontal area, *m* is the vehicle mass, $f_r$ is the rolling coefficient and *θ* is the road grade. $k_n$ (n=1, 2,…7) are initiated based on the different daily tasks probability and their contribution on the total Ah throughput of the battery. Initial value for $k_1$ is the driving task contribution on the *Ah*. This value for $k_2$, $k_3$, and $k_4$ which are related to the portion of each charging facility (L1, L2 and L3, respectively) is obtained based on a report from Idaho National Lab about the charging patterns of EVs in USA [28]. For $k_5$ to $k_7$, contribution of the solar integration, frequency regulation and peak shaving in the daily tasks' *Ah* is considered as the first guess.

The MCMC sampling starts with the defined initial values for all variables ($x^{(1)}$). In iteration *r*, a random proposal value ($x^{(P)}$) is offered based on the current value of the variable ($x^{(r)}$), for the next value of *x* as:

$$x^{(*)} \sim N(x^{(r)}, \sigma_t) \quad (15)$$

where, $\sigma_t$ is tuning SD which defines the range of proposal values and is selected based on the properties of each variable. It is normally smaller or equal to the SD of the prior distribution to guarantee a successful sampling. Note that the normal distribution in (15) is an example and the distributions for proposal values are same as each variable's prior distribution. Then, the full conditional is calculated for both the proposal and current values and Metropolis-Hastings sampling criteria is used to accept or reject the proposal value as (16).

$$R = \min(1, \frac{[x^{(P)}|\cdot]}{[x^{(r)}|\cdot]} \times \frac{[x^{(r)}|x^{(P)}]}{[x^{(P)}|x^{(r)}]}) \quad (16)$$

The proposal value is accepted as the next value of the variable in the chain ($x^{(r+1)}$) with probability $R$ and it is rejected and the current value is kept by probability 1-$R$. Applying this method to all variables and repeating it generates draws from the posteriors which can be used to estimate their distributions.

The last challenge in MCMC implementation is when the number of measurements in an observation is high (e.g., the acceleration data with resolution of 1 second in 30 minutes of driving). In that case, the multiplication of probabilities leads to a significantly small quantity that prevents the sampling criteria from accepting the proposal values. To solve this issue, we have reduced the number of data by taking their average value. However, to define the number of samples that are correlated and can be averaged, the autocorrelation of the data is calculated and the data points with an autocorrelation smaller than 0.2 are considered to be uncorrelated. The autocorrelation for observation data y can be calculated as (17).

$$A(lag) = \frac{\mathbf{E}[(y_s - \mu)(y_{s+lag} - \mu)]}{\sigma^2} \quad (17)$$

Where, $\mathbf{E}$ is expected value operator, s is sample index, $\mu$ and $\sigma$ are the average and SD of the data. As shown in Fig. 3 for a sample acceleration data, for the samples with index difference of 7 or greater, the autocorrelation is less than 0.2 (uncorrelated) and therefore, every 7 data points are used for reducing the number of data. This method is used for all input data.

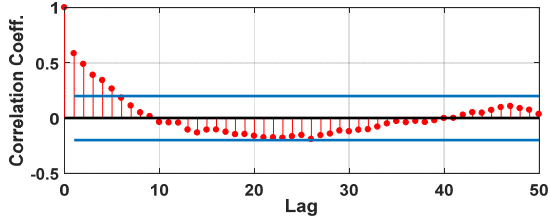

Fig. 3. Sample autocorrelation for acceleration data

IV. TRAINING DATA

To train the model and obtain the parameters, we have used three sets of experimental results on A123 ANR26650, 2.3 Ah LFP cells' capacity fade in different cycling conditions [29]. These data are shown in Fig. 4 which relates the capacity fade percentage to the total Ah throughput. The tests are performed different in temperature, SOC and C-rates. Table I shows the test condition for each of the cells. Note that these data are used only for model training and after tuning the model parameters, there is no need for the capacity measured data to estimate the battery health. So, in practice, a set of experimental (or manufacturer) data is needed to train the model, and after that, the model will estimate the battery health based on its performance condition without capacity measurement. Also, as the EV battery's thermal management system keeps the temperature in optimum operating range [30], [31], this model is trained using high temperature experimental data, the results are not valid for battery aging in extremely low temperatures. Low temperatures lead to Lithium plating due to the lower rate of lithium diffusion which reduces the cycleable lithium and causes capacity fade [32]. To use this model in low temperatures, it should be trained with proper data.

From these data, 70 data points are randomly selected to be used in training. The remaining 12 data points are kept for testing the model. Random selection of data is for the purpose of unbiased training and test results. The model obtains the measurements and initial values for the intermediate variables and generates chains of 10,000 samples for each variable using the mentioned MCMC sampling method. Fig. 5(a) shows the MCMC samples for one of the capacity fade data points which is drawn from posterior distribution. Based on this figure, the model is consistent in sampling and it does not wander around different values. These samples are used to fit distributions to the variables. Histogram and fitted gamma distribution of the samples in Fig. 5(a) are depicted in Fig. 5(b). Variation of fitted distribution's mean and SD for this data chain with 95% confidence interval are ±0.8 and ±0.03, respectively.

Table I. Test condition of LFP cells

| | Temperature (ºC) | C-rate | Average SOC | Max SOC | Min SOC |
|---|---|---|---|---|---|
| Cell A | 35 | 2.82 | 38.5 | 50 | 22.6 |
| Cell B | 23 | 3 | 41.9 | 54 | 32.5 |
| Cell C | 23 | 3.49 | 53.4 | 100 | 11.4 |

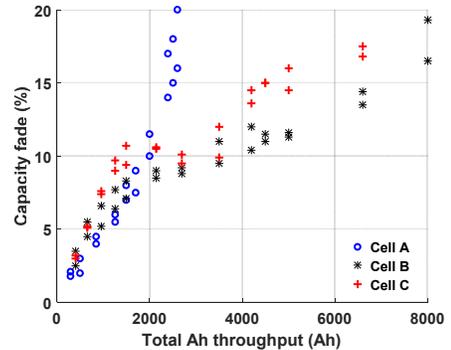

Fig. 4. Capacity fade data for three different test conditions

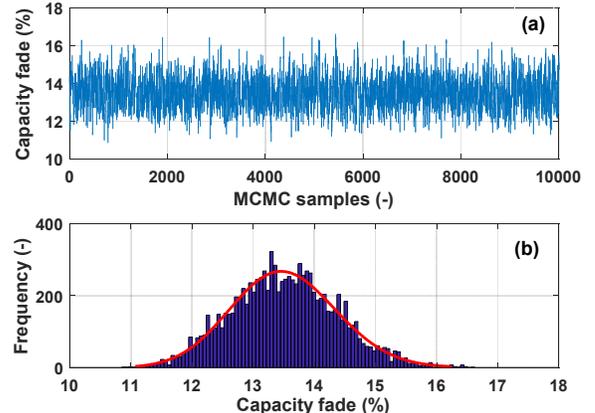

Fig. 5. (a) Chain of samples for a capacity fade data point and (b) Histogram and gamma distribution of the sample chain

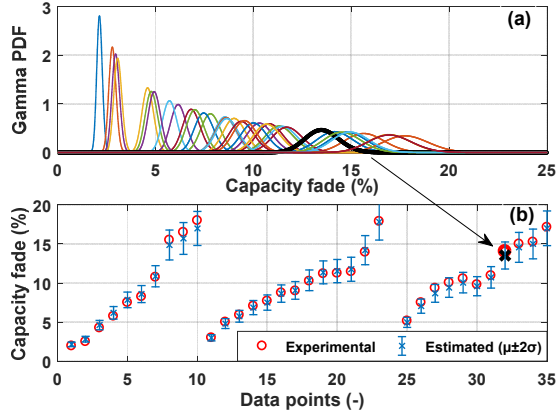

Fig. 6. Training set: (a) Capacity fade percentage gamma PDFs and (b) mean of PDFs compared to experimental data

The procedure is applied for all the variables and their fitted PDFs are obtained. The model's capacity fade estimation gamma distributions based on the measurements are depicted in Fig. 6(a). Each of these PDFs reveals the fact that the capacity fade percentage is not a fixed value and it varies between values with different probabilities to include the uncertainties of the measurements and the modeling process. For instance, the bold black plotted PDF in Fig. 6(a) shows variation between 11-17% capacity fade with average of 13.51% for two measurements of 13.6% and 14.5% and other uncertainties in the temperature, C-rate, and all other measurements.

To show that the results of the model's training are reliable, we have compared the mean value of PDFs in Fig. 6(a) to the mean of experimental measurements of the capacity fade (Fig. 6(b)). This figure also indicates the estimation dispersion in two SD interval ($\pm 2\sigma$). These results indicate that the model's training and parameters' tuning are successful.

Using the results of the training set, we have obtained the model parameters' distributions. These values are used in the model testing and case studies. It is important to note that an error parameter named residual error of capacity estimation $\varepsilon$ is introduced in the model to account for the estimation bias due to the lack of experimental data (Fig. 7). This parameter is used when estimating the capacity fade without experimental data.

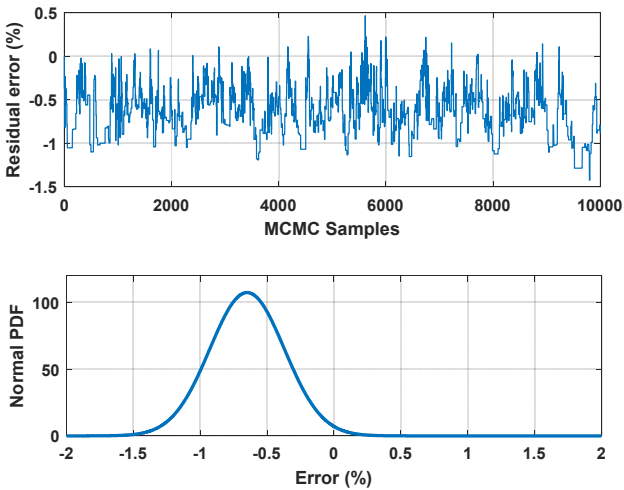

Fig. 7. Model bias parameter in capacity fade estimation

## V. MODEL EVALUATION WITH TEST SET

From the experimental data, 12 data points (different from the training set) are used to validate the model and test its estimation precision. Note, that these data points are for six observations with two measurements in each observation. The gamma distribution of the estimated capacity fade percentage for these observations are shown in Fig. 8(a) and the mean of these distributions are compared to the measurements in Fig. 8(b). Considering these figures, in most cases the estimated average is between two measured values and the model is successful in estimating the capacity fade percentage. To further validate the testing results of the model, 10 different training-test datasets are selected, and the training-test procedure is repeated. Fig. 9 shows the $R^2$ and percentage RMSD for estimation precision of the model in these 10 test datasets. Average $R^2$ of 0.94 and RMSD of 7% indicate acceptable precision of the model estimation.

Although these results show the model's high precision in 85/15% ratio of training/test data, it is important to explore the model's performance in different ratios, as well. Hence, to show the model's performance in different ratios of training/test sets and also time-series data selection (rather than random selection), we repeated training and test procedures for 4 ratios and 10 data sets (40 training/test runs in total). Ratios are 70/12 (85%), 60/22(70%), 50/32(55%) and 40/42 (40%). Also, 10 different training/test subsets of data pool for each ratio are selected in sequential manner. All data are sorted from lower to higher capacity fades as a result of longer cycling time. Fig. 10 shows the $R^2$ and percentage RMSD for these runs. Obviously, decreasing the number of training set, lowers the precision of model in the test results. Model's accuracy is acceptable for 85% and 70%, it is marginal in 55%, but not reliable in 40% training set percentage. Also, in lower percentage training sets, the variance of model precision is higher.

The developed model is a useful tool to estimate the EV battery aging in two important applications. One of the main challenges for the EV battery manufacturers is to define the battery warranty time/mileage which is a time consuming and costly process performed by controlled lab testing [33] that does not reflect the real-world experience of the battery. The developed model is a cost-effective and time-saving method to simulate the EV battery's real-world daily tasks accounting for as many as external effectual factors and define an optimum warranty system.

Another application is to implement the model in the onboard aging estimation. For this purpose, the model should be trained by manufacturer data for that specific battery chemistry and the parameters of the model should be tuned for the application specific (type of vehicle, duties, etc.) and environmental conditions. After proper training, the model will use the available online measurement data in the vehicle such as temperature, velocity, battery charging/discharging power profiles as its inputs and generate estimations on the battery capacity fade through probabilistic analysis of the data. Output of model can be used to inform the driver about the SOH of the battery. The model parameters should be retuned several times in the battery's lifespan to eliminate any accumulated estimation error over time which can be performed similar to the regular vehicle services.

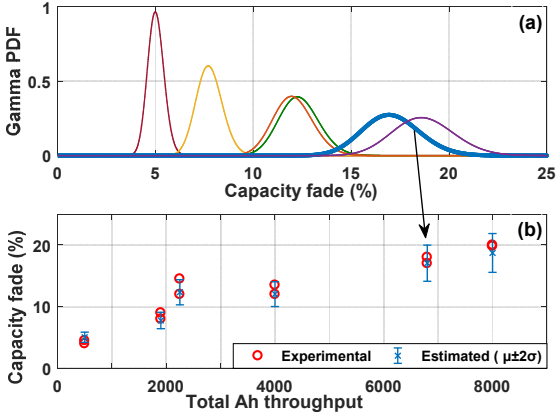

Fig. 8. Test set: (a) capacity fade estimation distributions and (b) comparison of estimated and measured capacity fade

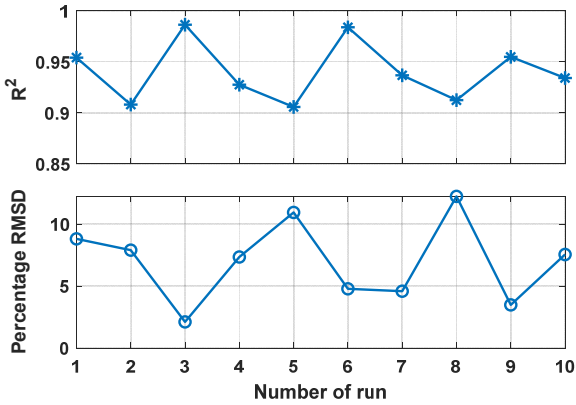

Fig. 9. $R^2$ and percentage RMSD in different training-test sets runs

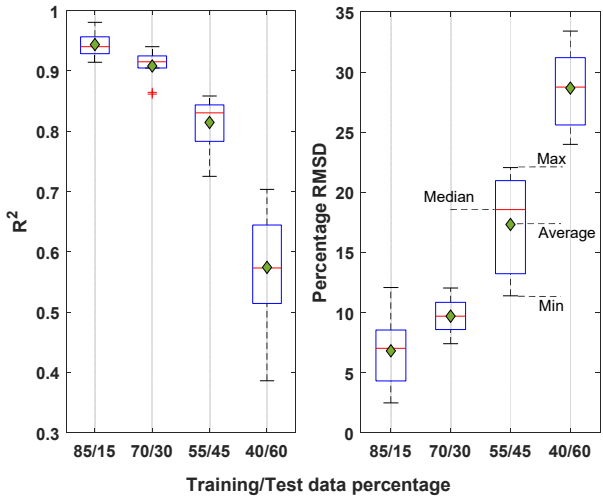

Fig. 10. Developed BN precision in different training/test data ratios

## VI. CASE STUDY: BN MODEL PERFORMANCE

In this section, we show how the developed BN model estimates the battery capacity fade in different cases with different provided data. Three cases are studied here. First, the effect of different drivers is explored by the model. The variables that are directly affected in this case are shown in Fig. 11 network with green color. The second case explores the impact of the frequency of grid services which directly changes contribution coefficients and the grid power (red part in Fig. 11). Last, the battery capacity fade in four different states of USA depending on their weather condition is simulated using daily temperature data (blue part in Fig. 11). Note that the battery aging tests are normally very time consuming even in the case of expedited aging studies and it is not possible to experimentally evaluate the battery aging behavior in the mentioned cases. Therefore, we are using BN model-based simulations to quantify the battery degradation under these scenarios.

### A. EV Daily Driving and Grid Services

In this case, it is considered that the EV is going to have its driving sessions in addition to grid services such as solar integration, frequency regulations and peak shaving, provided by it. Therefore, its battery cycling will include the driving, recharging and utility services. To simulate different drivers impact, recorded driving data of 45 drivers in Ann Arbor, MI [34] is used to simulate the daily driving pattern. For the recharging, based on an INL report [28], it is assumed that 38%, 55%, and 7% of all recharging events are performed by L1, L2, and L3 chargers, respectively.

Grid services are simulated by using recorded solar power data and PJM frequency regulations and peak shaving data. Note that this information is used to calculate the initial values of the contribution coefficients. The model is run for the capacity fade estimation after about 100,000 miles of driving which simulates almost 9 years of driving based on National Household Travel Survey (NHTS) [35].

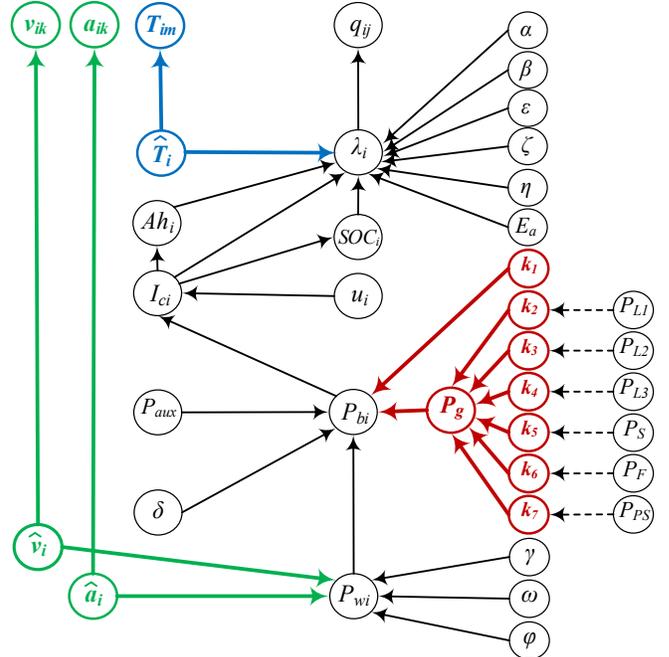

Fig. 11. Directly affected parts of BN network in driving data (green), grid services frequency (red) and different locations/weather (blue)

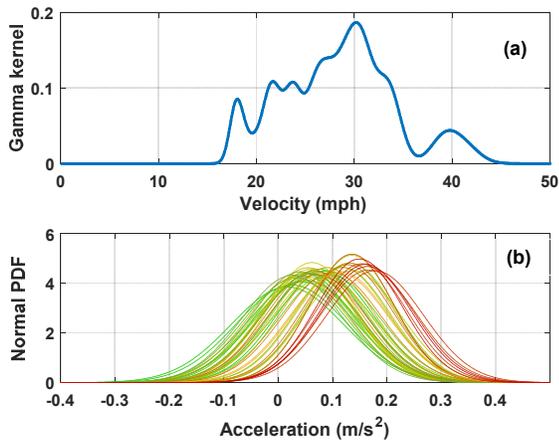

Fig. 12. (a) Velocity kernel and (b) acceleration PDFs

The model output for the driving cycles includes 45 independent distributions, with their kernel plotted in Fig. 12(a) to present the whole driving data distribution. This figure shows that the velocity data varies between 15 and 45 mph. Fig. 12(b) shows the acceleration PDFs for these driving cycles. The color code in this figure is based on the average absolute acceleration, (red to green color change refers to high to low values). We assume that the drivers with higher average acceleration (red curves) which means more frequent accelerations/decelerations during driving period are more aggressive drivers, while the gentle drivers have lower average acceleration (green). More information on how we concluded this can be found in [34].

The model's estimation for the efficiencies and the auxiliary power are shown in Fig. 13 (a) and (b). The battery power includes the effect of the charging events and grid services, as well. Therefore, the contribution of the driving and each of the recharging facilities in addition to grid services' contributions are estimated by the model and shown in Fig. 13(c) where, the driving's contribution is centered at 0.286 with higher values accounting for the regenerative braking and lower values referring to the auxiliary power. The L1, L2 and L3 chargers have contribution distributions, centered at 0.123, 0.18, and 0.021, respectively. Also, the contribution of solar integration, frequency regulation and peak shaving are 0.096, 0.093, and 0.151, respectively.

The model output for the SOC and Ah distributions are depicted in Fig. 14(a) and (b). These distributions show the effect of all EV tasks, including grid services. Also, it indicates that the gentle drivers SOC distributions have higher mean values, while there is no significant difference on the distributions' variance. However, they have lower total Ah throughput mean values with lower variances which shows that the model estimates the higher Ah distributions with more uncertainties. It is important to note that the SOC and Ah have direct relationship with capacity fade and therefore, based on the trend of these distributions, these two factors are in race.

The most important result of the model is capacity fade percentage estimation which is shown in Fig. 15. Considering Fig. 15(a), the capacity fade PDFs have higher mean value for the aggressive drivers. The SD of distributions are similar, varying between 1.2 and 1.4. From red to green, the SD is decreasing, which can be the effect of Ah distributions' SD. It is important to note that the standard deviation of an estimated distribution is dependent to the amount of uncertainties in its parent nodes. Therefore, to decrease the uncertainty in capacity fade estimation, the uncertainties in the measurements and process model should be decreased. More accurate measurements and more precise process model will improve the estimation precision. Comparing the effects of Ah and SOC in Fig. 15(b) indicates that the Ah is more dominant factor in capacity fade, as the drivers with higher Ah have higher capacity fade at the end of driving period, although their SOC is lower. Considering the color change (refers to average acceleration) in these figures' data, there are outlier points which indicates that the capacity fade has not linear relationship with the acceleration profile. However, the general trend shows strong relationship between aging and driving aggressiveness.

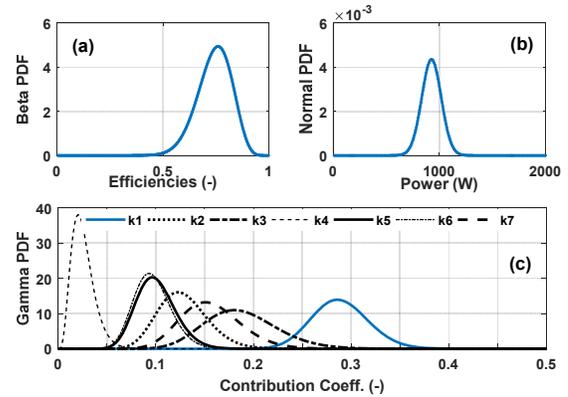

Fig. 13. PDFs of (a) EV power train efficiencies, (b) Auxiliary power and (c) contribution coefficients of driving and recharging events

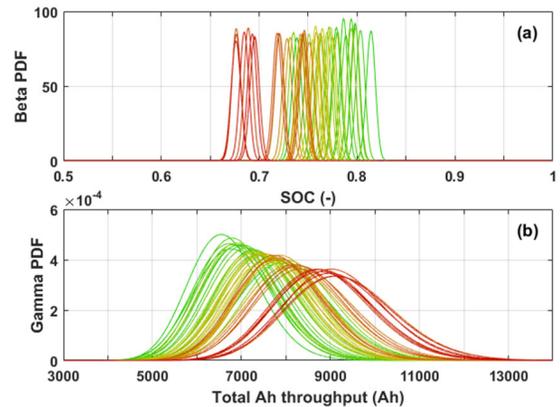

Fig. 14. (a) Beta PDFs of SOC and (b) gamma PDFs of Ah for all drivers

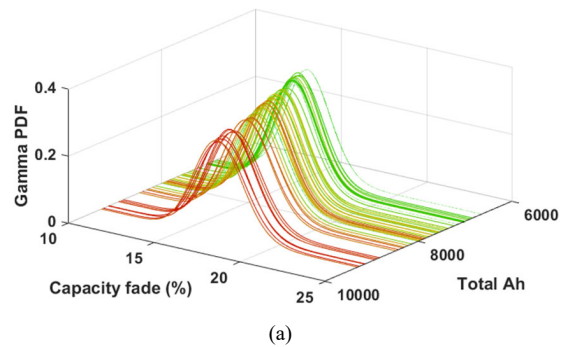

(a)

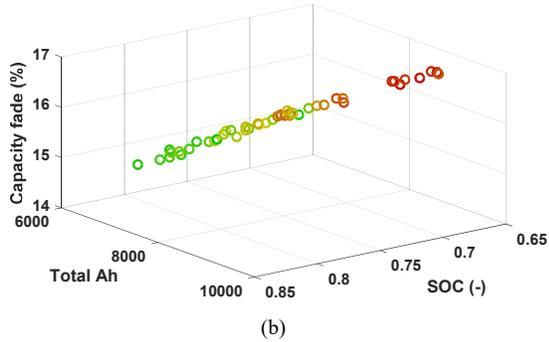

(b)

Fig. 15. (a) Capacity fade PDFs and (b) mean of capacity fade PDFs respect to Ah and SOC means

### B. Effect of grid services' contributions

As, there is no defined pattern for the grid services and availability of the service, the infrastructures and the willingness of the vehicle owner varies it significantly, we collected the potential daily tasks of an EV battery in Fig. 16 which includes three driving sessions, charging by solar panels, frequency regulation, peak shaving and standard charging. Note that this pattern is not necessarily followed every day by every vehicle and it is only to show the variety of tasks. To show the intensity of these tasks on the battery cycling, Table II presents the battery cell level performance results for each of these tasks which directly affect the capacity fade. Table II shows that although the main performance of the battery is dedicated to the driving which is clear from the total Ah and DOD comparison, sum of all grid services can occupy a significant portion of daily performance. Among these grid services, peak shaving has deeper cycling effects on the battery. Note that the grid services are considered to be performed by the standard L2 charging station. Also, depending on the different daily patterns, standard charging can happen in different day times with different standard chargers.

To explore the effect of each service on the battery capacity fade, a daily scenario is considered with driving, only one grid service and recharging. Two hours of service is considered for the frequency regulations and peak shaving, while solar integration is simulated by 3 hours charging through solar panel output power. Fig. 17 depicts the effect of individual grid service on the cycling condition and capacity fade of the battery.

"Driving only" is the base case for comparison. Daily frequency regulations have almost no effect on the average cycling SOC of the battery due to its negligible DOD, however it decreases the average C-rate of the whole daily cycle. It also increases the total Ah throughput equivalent to 115 full cycles. All these changes cause 7% more capacity fade in whole life of the battery. Solar charging happens in the middle of day and therefore increases the daily average SOC of the battery, while decreasing the average C-rate. It does not change the total Ah throughput, and its effect on the capacity fade is negligible. Peak shaving has the most dominant effect on the battery performance and as a result capacity fade percentage. This service increases the C-rate while decreasing the average SOC. Also, because it is discharging, it significantly affects the total

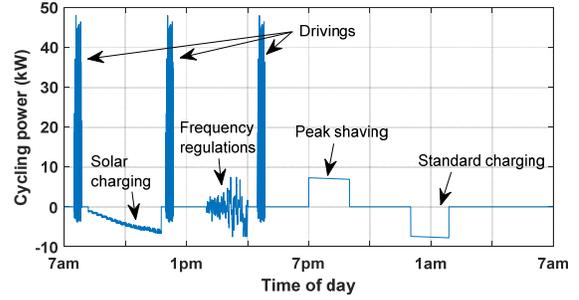

Fig. 16. Potential daily tasks of an EV battery

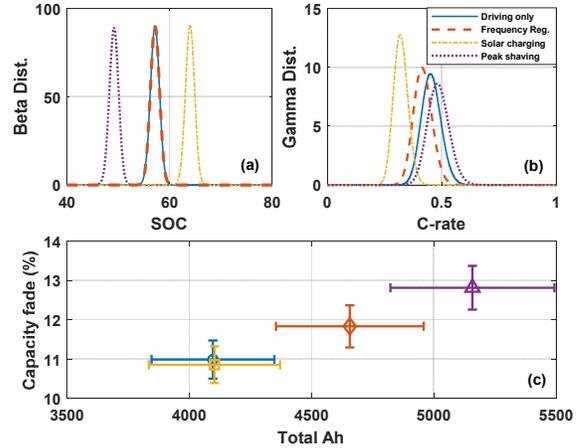

Fig. 17. Impact of grid services on the battery cycling and capacity fade percentage: (a) SOC, (b) C-rates and (c) capacity fade percentage vs. total Ah throughput (μ±σ)

Table II. Battery cell level performance result in potential daily tasks

|  | Total Ah | C-rate | Average DOD (%) |
|---|---|---|---|
| Driving | 0.76 | 0.2 | 31.5 |
| Solar charging | 0.27 | 0.1 | 11 |
| Frequency Reg. | 0.26 | 0.13 | 0 |
| Peak shaving | 0.43 | 0.5 | 18 |
| Standard charging | 0.48-0.9 | 0.1,0.5,5.5 | 20-38 |

Ah throughput by increasing the required charging at night (225 full cycles in total). Consequently, it leads to extra 15% capacity fade over life of the battery. Increasing the number of services and combining them into the daily tasks of the EV battery boosts the risk of higher degradation. To evaluate the most extreme case with highest impact on the battery, we developed a scenario which performs all mentioned services every day and compared it to the case of no grid services. We also developed several cases in between as once a month, every week and every other day grid services frequency. Running the model for these patterns changes the contribution coefficient's estimation as shown in Fig. 18. In this figure, $k_1$ (driving) is shown in trace (a), $k_2$ to $k_3$ (charging) and $k_5$ to $k_7$ (grid services) are aggregated and shown in trace (b) and (c), respectively.

It is observed that if the frequency of grid services events increases, the portion of driving and recharging in the total battery Ah decreases, although the total Ah increases significantly by adding more grid services events (Fig. 19 (a)). This growth in Ah increases the capacity fade percentage accordingly, as indicated in Fig. 19(b).

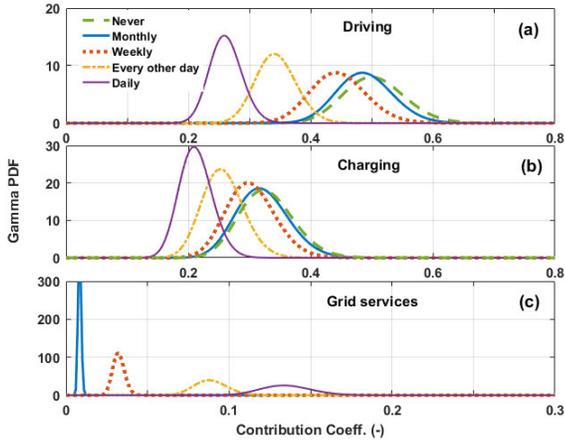

Fig. 18. Contribution Coefficients PDFs in different frequency of grid services events

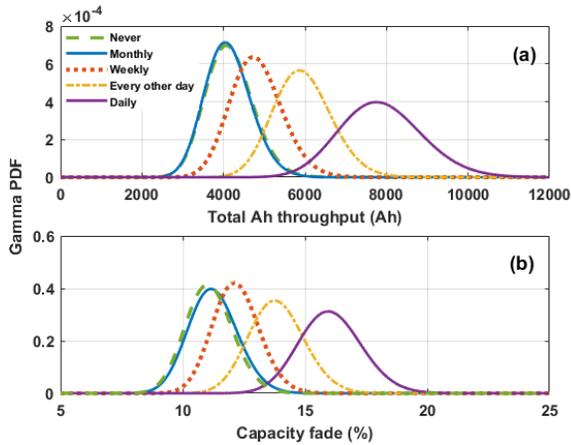

Fig. 19. (a) Total Ah and (b) capacity fade PDFs for different frequency of grid services events

Adding different number of these events changes the battery power, C-rate and SOC distributions as well. It is important to note that having three grid services every day is a very extreme case which does not happen in normal daily patterns. However, it is to show what is the effect of extreme grid-side cycling on the EV battery.

### C. Impact of location/weather change

If the EV is driven in different climates, the capacity fade will change based on the temperature pattern. To explore the effect of temperature change, we have extracted the daily temperature data for four US cities with different climates (Phoenix AZ, Ann Arbor MI, Miami FL, and Portland OR) in 2016 and run the model for these temperature observations. To show the performance of model in the temperature distribution estimation, temperature data for Portland and output of model for its estimation is depicted in Fig. 20. In this result, the probability distribution type for temperature is defined as Normal distribution and the parameters for this distribution are tuned by the Bayesian estimator.

The normal distribution of the annual temperature in these cities are shown in Fig. 21 (a). These temperature patterns lead to different capacity fade estimations as plotted in Fig. 21 (b). From the figure, there is no significant difference between Phoenix and Miami, as their average temperature is similar, however the small difference can be caused by the higher range of temperature change in Phoenix, which offers harsher temperature conditions for the battery. Ann Arbor has a wide annual temperature range because of its cold winters, while Portland's temperature distribution has smaller SD. Lowest capacity fade happens in Portland and highest is in Phoenix.

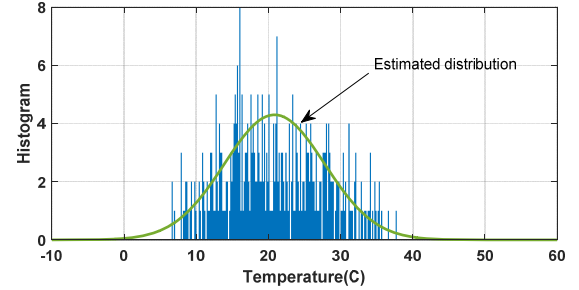

Fig. 20. Sample temperature data and distribution estimation by the BN

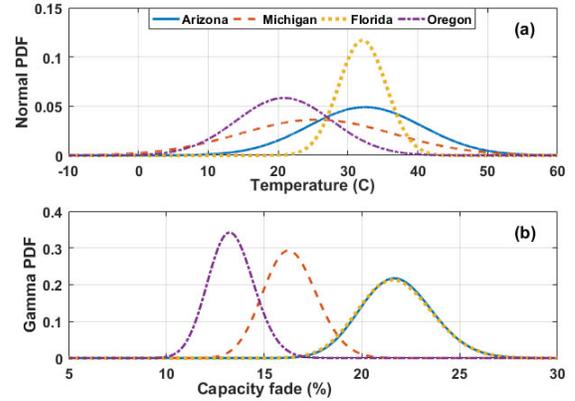

Fig. 21. (a) Temperature and (b) capacity fade distributions for different climates

### VII. CONCLUSION

The battery capacity fade modeling and estimation has uncertainties due to the probabilistic nature of its effectual factors. Therefore, deterministic models cannot completely explain the capacity fade process. Bayesian models are proved to successfully map the causality in probabilistic processes in ecological systems and the battery aging phenomena presents a similar problem. Therefore, in this paper, we proposed a probabilistic analysis using a hierarchical Bayesian model for the EV battery capacity fade estimation to include the uncertainties in the variables' measurements and modeling process. The Bayesian network for this purpose is developed including all random variables and Metropolis-Hastings MCMC sampling method is used to calculate the posterior distributions. This model estimates the capacity fade with high precision while accounting for multi-layered operational and environmental factors and therefore it can help the manufacturers in more accurate warranty definition and the costumers and drivers through onboard battery health estimations. The model is trained and tested by a set of experimental data. Test results show that the model is successful in estimating the capacity fade percentage with 94% accuracy in 85/15% training/test data ratio. However, the model's accuracy drops in lower ratios of the training set. Also, three case studies as the effect of different driving styles, different grid services tasks, and different temperature profiles are explored by the model to show its performance in different input data. The outputs of model for all the variables are probability distributions which reflect their inherited uncertainties. Here are some of the highlighted results:

- In the case of daily driving, recharging and grid services, the battery's cycling varies from 1250 to 2750 (full cycle equivalent) for gentle to aggressive drivers. This leads to 13-20% capacity fade.
- Normal grid services pattern (one service daily) does not affect the battery life significantly. Service type also defines the the battery performance. Peak shaving has highest impact with maximum 15% extra capacity fade.
- Extreme temperatures have a significant effect on the capacity fade. In moderate climate of Oregon, the battery has 13% capacity fade on average, while is Arizona and Florida, this number is 22%. Michigan is in between with 16.5%.

The merit of developed model is that it is applicable for any kind of experimental/measured inputs, if it is trained by the prior data and it can be combined with other Bayesian models. Note that the results of this study are valid only in the range of temperature that the model is trained. For more comprehensive study, the model should be trained in higher range of temperatures including low temperature experimental results. Future work will focus on developing more detailed Bayesian model for whole life of a battery from EV to stationary applications.

ACKNOWLEDGMENT

Authors would like to acknowledge Dr. Kuilin Zhang from Michigan Technological University for providing the driving data.